# Physics in Carnacki's investigations

A. Sparavigna

Dipartimento di Fisica, Politecnico di Torino

C.so Duca degli Abruzzi 24, 10129 Torino, Italy

In the stories of Carnacki, created by the English writer William H. Hodgson and written between 1910 and 1912, we find an interesting mixture of science and fantasy. In spite of the fact that Carnacki is a ghost finder, who investigates in an environment where supernatural is acting, we see the character using technology and sometimes discussing of physics phenomena such as X-rays. The Carnacki's stories can be seen as mirrors of the physics and technology knowledge at the beginning of 20th century, in the society outside scientific circles.

**Introduction.**

Thomas Carnacki is a fictional detective, investigating facts with a possible supernatural origin, created by William Hope Hodgson (1877-1918), an English writer. Carnacki the Ghost Finder is the protagonist of short stories [1], published between 1910 and 1913 and posthumously. As the well-know detective Sherlock Holmes, Carnacki lives as a bachelor in Chelsea, London. Following the style of Arthur Conan Doyle, where the adventures of Sherlock Holmes are reported by Watson, the investigations of Carnacki are told from Dodgson, one of Carnacki's four friends, in a first-person perspective. "In response to Carnacki's usual card of invitation to have dinner and listen to a story", the five friends meet at 427, Cheyne Walk, and after dinner they are engaged in the report of supernatural adventures.

When Carnacki is called for help when haunting is suspected, he uses tools from tradition such as chalk, water and candles drawing pentacles in old rituals to protect himself. Surprisingly, Carnacky is inventing an electric pentacle for protection, using the new technologies of the early 20$^{th}$ century. The sensitivity of photographic plates is also used to reveal the presence of invisible things. Unlike the Sherlock Holmes's character, the Carnacki's scientific background is never disclosed in detail by the writer, but he knows physics and chemistry for sure. Dr.Hesselius [2], the scientist disposed to supernatural, protagonist of short stories by Joseph Sheridan Le Fanu, probably inspired the character of Carnacki too. And Hodgson's works probably influenced later writers, among whom Seabury Quinn who developed the supernatural detective Jules de Grandin [3].

Being the Carnacki's character not dogmatic, he always uses evidences to reach a conclusion. In some stories, the haunting is real, while in others the problem is not clearly solved and the reader is never sure if ghosts exist. The nature of Hodgson's creation is that of a generic gentleman of England, never exceeding the story frame. What is fascinating in the Ghost Finder's stories is the environment, where supernatural is possible but not clearly sure. Moreover, the presence of science in an intriguing mixture with fantasy could be interesting for the reader. Carnacki's stories can be seen as mirrors about the knowledge of physics and technology at the beginning of 20th century of people outside the scientific academies.

**Electricity, optics and vibrations.**

As told before, Hodgson was able to create a true suspense between haunting and real facts. In "The Gateway of the Monster", 1910 [4], Carnacki is engaged by a gentleman to investigate a noisy spirit spending each night slamming the door of a room in the manor house. This is the first story where Carnacki uses the electric pentacle to protect himself while spending the nights in that room. He succeeds trapping the haunting entity in this pentacle. In "The House Among the Laurels", 1910 [5], a deserted mansion displays signs of haunting. Carnacki realises that he had been played for fool. He discovered, using a camera and studying photographic plates, that a criminal gang is actually living hidden in the mansion, taking advantages of frightening people with haunting tricks.

In "The Thing Invisible", 1912 [6], Carnacki places the camera ready to photograph any mysterious phenomena in the dark of a chapel attached to an Edwardian manor house. In the chapel, a peculiar old dagger is kept: after the nightfall, the haunted dagger attacks any enemy who should dare to venture into the Chapel. In the dark of the night, Carnacki hears mysterious noises and when he approaches the altar, the dagger nearly kills him. The photographic evidence shows the truth to Carnacki: subtle differences between photographic plates reveal the mechanism of an ancient trap armed to guard the altar.

In "The Hog", published in 1947, Carnacki faces a disturbing giant hog spirit that is trying to enter the real world from horrifying nightmares. He is equipped with a new variant on the electric pentacle, developed to record the nightmares when connected to the head of a dreamer. In this story, Carnacki reproduces the nightmares recorded on a paper band as sounds. In "The Haunted Jarvee", published in 1929, we find a device to create vibrations. In the story, Carnacki decides to go aboard the Jarvee, antique sailing ship, to investigate a possible haunting. In the night, shadows are appearing from the surface of the sea, running as waves towards the ship. Close to the Jarvee, they disappear from sight. After a while, furious storm arises which elapses all the night. Carnacki guesses that the phenomenon is caused by vibrations and develops a device to emit repellent vibrations. After first good results, the machine is not able to suppress more furious storms, and at the end Carnacki and the crew are forced to abandon the Jarvee that sinks to the bottom of the ocean. Dodgson asks what caused the haunting and Carnacki explains his theory of "focuses", saying that the Jarvee was a sort of focal point for vibrations. Replied Carnacki, 'in my opinion she was a focus. That is a technical term which I can best explain by saying that she possessed the "attractive vibration" that is the power to draw to her any psychic waves in the vicinity, much in the way of a medium. The way in which the "vibration" is acquired - to use a technical term again - is, of course, purely a matter for supposition.' [7]

The explanation has its frame in the theories about spiritual vibrations, connected to Spiritualism, where mediums are focussing elements. Spiritualism was a religious movement that began in the United States and was prominent between 1840s and 1920s in English-speaking countries. Its distinguishing belief is that spirits of dead people live a higher level of existence, but they are able to become spiritual guides when Mediums establish a relationship. In the case of English naturalist Alfred R. Wallace [8] and of Sir Willian Croockes, chemist and physicist, Spiritualism was a matter of science, not only a religious belief. We can imagine that Hodgson was aware about the book of 1904 by Crookes [9]. To tell the true, the Jarvee looks like a receiver of a radio station, the first experiment of a transatlantic radio communication was in 1901 by Gulgiemo Marconi. "The Hog" and "The Haunted Jarvee" were published after the dead of the author in 1918. Here we prefer to consider just the stories published between 1910-1913, at the beginning of the century and discuss links to science that we can find. Let us just tell that in "the Haunted Jarvee" we can find a very impressive description of St. Elmo's Fires, that is the electrical phenomenon in which luminous plasma is created by a corona discharge at the ship's masts, due to atmospheric electric field.

**Glowing gases.**

The "electric pentacle" is a device invented by Hodgson for his character to protect himself from supernatural forces, ghosts or monsters. The device consists of a series of wires and glowing vacuum tubes, a realistic portrait of electric technology in early 20$^{th}$ century. Hodgson describes the elaborate set-up, which Carnacki prepares before spending the night in a haunted room, to complete more traditional protective arrangements. He developed this second technological defence, after becoming aware of the "Professor Garder's 'Experiments with a Medium.' When they surrounded the Medium with a current, in vacuum, he lost his power - almost as if it cut him off from the Immaterial. That made me think a lot; and that is how I came to make the Electric Pentacle, which is a most marvelous 'Defense' against certain manifestations. I used the shape of the defensive star for this protection, because I have, personally, no doubt at all but that there is some extraordinary virtue in the old magic figure. Curious thing for a Twentieth Century man to admit, is it not? … I turned-to now to fit the Electric Pentacle, setting it so that each of its 'points' and 'vales' coincided exactly with the 'points' and 'vales' of the drawn pentagram upon the floor. Then I connected up the battery, and the next instant the pale blue glare from the intertwining vacuum tubes shone out" [4]. The same device used in "The Gateway of the Monster" is described in "The House Among the Laurels": the pentacle, when connected up the batteries, shines a weak blue glare from vacuum tubes.

Carnacki had the idea of an electric pentacle after reading the Garder's experiments with a medium. Of course, such a publication by Garder as author does not exist, but we can suppose that such experiments with glowing tubes were performed at the beginning of 20$^{th}$ century and reported in magazines. Let us note that Sir Crookes dealt with several famous spiritual mediums and this could have inspired the figure of a scientist investigating supernatural fact. Sir Crookes actually made the first experimental observations of new phenomena connected with electrical discharge in low-pressure gases in 1898 [10], the first steps of plasma science, and the word "plasma" may possibly had a previous meaning related to Spiritualism [11].

We can imagine Carnacki's electric pentacle made as a neon luminous tube sign, containing neon or other inert gases at a low pressure. A high voltage applied to electrodes makes the gas glow brightly. The neon sign is an evolution of the earlier glass tubes for demonstrating the principles of electrical discharge in gases. But were neon signs existing in 1910? The Nikola Tesla's neon lamp signs were firstly displayed at the 1893 Chicago World's Fair [12,13]. Later in 1897, D. McFarlan Moore gave a demonstration of his results in developing glowing lamps [14], producing "the daylight in a tube" and disclosing a lighting system never seen before. The development of neon signs is credited to Georges Claude and the first public display of a neon sign was in December of 1910 at the Paris Expo and first commercial signs were sold in 1912 [13]. Hodgson seems to know this new technological application of fundamental studies on glowing discharges in low-pressure gases. But in fact he knows much more.

**Charnacki and the "lightless" photography.**

As we have reported before, Carnacki uses photography in almost all his adventures. As in "The Horse of the Invisible", 1910 [15], Carnacki performs experimental photographs of a subject and her/his surrounding, because "sometimes the camera sees things that would seem very strange to normal human eyesight… I asked Miss Hisgins to join me in my experiments. She seemed glad to do this and I spent several hours with her, wandering all over the house, from room to room and whenever the impulse came I took a flashlight of her and the room or corridor in which we chanced to be at the moment." [15] Carnacki works in the hope that sensitivity of photographic plates is able to reveal an invisible presence near the subject. And this is more or less a procedure, commonly believed to reveal supernatural spirits.

When Hodgson was writing the adventures of Carnacki, long time had passed from the first permanent photograph image produced in 1826 by Nicéphore Niépce, French inventor. The "Boulevard du Temple", taken by Louis Daguerre in 1838 or 1839, can be considered the first known photograph of a person. The subject was a boulevard but since exposure time was over ten minutes, moving people and traffic do not appear. Just a man was recorded in a corner of the image frame, because he stood still while getting his boots polished. In 1839 Daguerre announced the development of the process for daguerreotypes. In fact, in 1832, Hercules Florence had already created a very similar process, naming it Photographie [16].

The modern photographic process came about from a series of refinements and improvements: let us just remember George Eastman. In July of 1888 Eastman's Kodak camera went on the market turning the image recording in a popular resource, and finally in 1901 reached the mass-market with the Kodak Brownie camera. Hodgson was really fond of photography and often illustrated his lectures with coloured slides (see Appendix on Hodgson's life).

We can consider the photography as a well-known technology in 1910 and then quite natural for the Hodgson's character to use it, but Carnacki is surprising. In the story "The Thing Invisible" of 1912, he claims his interest in the "lightless photography". "I set to work immediately to develop, not the plate I had exposed, but the one that had been in the camera during all the time of waiting in the darkness. You see, the lens had been uncapped all that while … You all know something of my experiments in lightless photography, that is "lightless" so far as our human eyes are capable of appreciating light? It was X-ray work that started me in that direction, and now I had vague and indefinite hopes that, if anything immaterial had been moving in the chapel the camera might have recorded it." This quotation about X-rays is rather interesting. Let us report a short history of X-rays where we shall see the connection with photography, in particular with the "lightless" photography

**From Crookes to X-rays.**

Let us go back to discuss the electric currents in gases at low pressure. In 1879, during a lecture delivered to the British Association for the Advancement of Science at Sheffield, William Crookes discussed his experiments with what is called now the Crookes tube, a glass cylinder mostly evacuated, containing electrodes for discharges of a high voltage electric current. He reported that the walls of the tube are emitting a dull blue colour and that when unexposed photographic plates are placed near the tube, some of them show shadows. Low-level X-rays produce the blue glow when the cathode rays struck the glass.

In his paper of 1891 [17], Crookes refers to Puluj. Ten years before, Ivan Puluj invented a device for producing what he called "cold light" and used it for a kind of X-ray photography. The X-ray emitting device became known as the Puluj lamp and produced for a period [18]. Puluj published his results in a paper entitled "Luminous Electrical Matter and the Fourth State of Matter" in the Notes of the Austrian Imperial Academy of Sciences (1880-1883) [19]. Let us note that this is the terminology of plasma physics. According to [18], the paper was rather obscure and this fact could explain the disappointment of Crookes who disagreed with Puluj on the nature of radiant matter. We guess that Hodgson, when he studied in Liverpool to receive the mate certificate, had the possibility to discuss about developments of physics and technology. From his studies and readings in popular-science magazines or more specific journals, he probably gained some knowledge of problems concerning X-rays and radiant matter.

In that period there was a strong scientific debate about the radiant state of matter. As Faraday proposed, the matter was classified in four states: solid, liquid, gas and radiant [11]. Researches on the last form of matter started with the studies of Heinrich Geissler (1814-1879): the new discovered phenomena, different from anything previously observed, persuaded the scientists that they were facing with matter in a different state [20]. Crookes took again the term "radiant matter" coined by Faraday: in Ref.17, it is clearly stated than the radiant matter is connected with residual molecules of gas in the tube.

In April 1887, Nikola Tesla began to investigate X-rays using high voltages and tubes of his own design, as well as Crookes tubes. After several years of experiments, in 1897 during a lecture before the New York Academy of Sciences [21], Tesla stated the method of construction and safe operation of X-ray equipments, due to the biological hazards associated with X-ray exposure [22]. Wilhelm C. Röntgen was the first to perform a systematic study of X-rays and recognise them as a new kind and unknown type of radiation, referring to it as "X". For this discovery, Röntgen receives the first Nobel Prize in Physics. In 1896. Rontgen gave a demonstration of the medical use of X-ray when he saw a picture of his wife's hand, a radiograph, on a photographic plate. In the same period, Thomas Edison invented the fluoroscope that immediately became a standard device for medical X-ray exam [23]. The connection of X-ray and photography, as "lightless photography", is quite evident.

Let us remember Fernando Sanford too and his experiences with photography [24], because he did a lot of experiments with X-rays. Previously in 1893, in a letter to The Physical Review [25] he described the discovery of the "electric photography". It is a contact-print photography, and if we see the image of a coin obtained by Sanford in [25], we immediately acknowledge the similarity with Kirlian images. The Kirlian photography refers to a contact photography associated with high voltage, named after Semyon Kirlian, who discovered it in 1939. If an object on a photographic plate is connected to a high voltage source, the corona discharges created by the field at the object edges produce an image on the plate. Sanford wrote also an article entitled "Without Lens or Light, Photographs Taken With Plate and Object in Darkness", published in the San Francisco Examiner [26].

Prior to the 20$^{th}$ century and for a short while after, X-rays were generated in cold cathode tubes. The tubes contained a small quantity of gas, to allow a current to flow. In 1904, John A. Fleming invented the thermionic diode valve, where heated cathode allows the current in vacuum due to thermionic emission of carriers. The principle was quickly applied to X-ray tubes: we could consider this as starting point of the modern X-ray investigation. Two years later, Charles Barkla (Nobel Laureate in 1917) discovered that X-rays could be scattered by gases, and that each element had a characteristic X-ray. Max von Laue, Paul Knipping and Walter Friedrich observed for the first time the diffraction of X-rays by crystals in 1912. This discovery gave birth to the field of X-ray crystallography. Many other scientists concurred in the development of X-ray technology and its medical use, in some cases injured by the dangerous radiation.

**Conclusions**

During a short discussion on the Hodgson's knowledge of physics, which we deduced from Carnacki's stories, we have also reported a small part of the strong debate about four states of matter at the end of 19<sup>th</sup> century. We can imagine echoes on journals and popular magazines and, at the same time, the spreading all over the world of new electric technologies. Concerning Hogdson, he had a solid scientific background from schools probably, reinforced by a true love for science, demonstrated by his experiences in photography of natural phenomena. His main feature can be described by a sentence Carnacki told, "I ask questions, and keep my eyes open".

**Appendix: Hodgson's life.**

Hodgson was born in 1877 in Blackmore End, Essex [27]. At the age of thirteen, Hodgson ran away from school in an effort to become a sailor. Caught, he returned to his family but eventually received his father's permission to begin an apprenticeship as a cabin boy in 1891. After his apprenticeship ended in 1895, Hodgson studied in Liverpool and received the mate certificate. He was then sailor for several years. At sea, Hodgson experienced bullying [28]; the theme of bullying and revenge appeared frequently in his sea stories.

The motivation of self-defence induced him in training his body. At sea, in addition to his exercises with weights and punching bag, Hodgson practised photography, taking among others photographs of atmospheric phenomena such as cyclones, lightning and aurora borealis. During his period as sailor, he was awarded the Royal Humane Society medal for heroism. Back from sailing, he opened in 1899 a school of physical culture, in Blackburn, England, offering personal training.

In this period he began writing articles on physical culture, featuring photographs of himself while demonstrates exercises. After these articles, Hodgson turned his attention to fiction, publishing his first short stories. Moreover he began to give paid lectures illustrated with coloured slides, about his life as sailor. He wrote poems too, many posthumously published by his widow. Sea stories were sold to American and British magazines: in 1907 he published "The Voice in the Night", a horror sea story, and "The House on the Borderland". In 1909, Hodgson published "Out of the Storm", horror story about "the death-side of the sea", and the novel "The Ghost Pirates". Hodgson's most famous short sea story is "The Voice in the Night", adapted for film twice.

In spite of the fact that these novels received a critical success, the author remained relatively poor. To increase his income, he began working on the first of his recurring characters, Carnacki, featured in his most famous stories. The first of these, "The Gateway of the Monster", was published in 1910 in The Idler. Another recurring character is Captain Gault, a sea captain for hire.

In 1912, Hodgson married Betty Farnworth, a staff member of a magazine. After a honeymoon in the south of France, they took up residence there. Returned with his wife to England, he joined the London's Officer's Training Corps and received the commission as Lieutenant in the Royal Artillery. In 1916, he suffered injuries from an accident, and discharged from the army, he returned to writing. Sufficiently recovered, he was re-enlisted. He wrote articles and stories reporting his experience during the war. An artillery shell at Ypres killed him in April of 1918. His widow worked to keep his books in print and, after her death in 1943, Hodgson's sister took over his literary legacy.